\documentstyle[graphics,aps]{revtex}
\begin{document}
\draft
\title{Gravo-inertial field theory}
\author{Doriana Kiekh\"oven} \address{Leibniz-Akademie und -Arbeitskreis Berlin e. V.,
Am Graben 3, D-15732 Eichwalde, Germany}
\date{February 26, 2000}
\maketitle
\begin{abstract}
 Inertia is defined axiomatically. The gravitational field is caused by
the flow of intergalactic masses. Origin of space and time are connected
with fields. The cosmos is bounded by inertia and gravitation, which is
the sequence of existence of two fields, the inertial field and the
gravic  field as a vortex field.Gravic and inertial field combine to a
unit, the gravo-inertial field. The separation in gravic and inertial
components depend on the coordinate system of motion.
\end{abstract}
\section{Historical}

 Well known models like "steady state theory",
Einstein cosmos, Friedman cosmos are although the last expanding,
"static" models, strong determined by geometry or rigid
attemptions. The gravitational field is given, but no  created.
The deforming of space and time is necessary because given at the
beginning. This may be the analog and appropriate way with a
mathematical method describe a physical problem.  We tried to
explain background  with considering the results of the mechanics
of H. Hertz, V. Fock and the the epistemological considerations of
G. W. Leibniz We will outline in condensed summary our
investigations about gravitation and inertia.
\section{Axiomatic}
\begin{enumerate}
\item Law of inertia mass
\item Conservation law
\item Special theory of relativity
\item Continuity principle
\item Creation of time and space by field and vice versa
\item Gravo-inertial field equations
\end{enumerate}
The priciples 1, 2  and  3 are well proved. We outline 5 and 6
considering 3. Starting point are the existence of mass and inertia.
Mass is existent by space (extent) and time (motion).M otion and density
change is influenced by two fields,  first ($\mathbf\vec{K}$)  arised by inertia,
second ($\mathbf\vec{G}$) by gravitation, the later conditioned by motion and density
change (enlargement).
\section{Experimental}
The cosmos is at present expanding with definit velocity and
accelleration. The equality of gravitational and inertial mass is
assumed proved. The background radiation is a characteristic of an
earlier state. The present state is a result of development. Gravitation
 can assumed as an effect of higher order. There is a strange numeric
relation between electricity and gravitation of the electron. The
repulsion of two electrons is 1040  times of the attraction as the
result of gravitation, the rules are similar.
\section{Gravo-inertial field equations}
The gravo-inertial field equations can be written in MKS-System as
\begin{equation}
\nabla\cdot\mathbf\vec{K}=-\frac{\partial\varrho}{\partial
t}+\cdot\cdot\cdot\text{+higher
orders}
\end{equation}
\begin{equation}
\nabla\cdot\mathbf\vec{G}=0
\end{equation}
\begin{equation}
c^2\nabla\times\mathbf\vec{G}=-{\frac{\partial\mathbf\vec{j}}{\partial t}}+\cdot\cdot\cdot\text{+higher
orders}
\end{equation}
\begin{equation}
\nabla\times\mathbf\vec{K}=-\frac{\partial\mathbf\vec{G}}{\partial t}
\end{equation}\\
where \\$\vec{j}$ [kg/$m^2$s] mass flow (in an isotropic world
radial to the reference system)\\
\\ $\rho$  mass density [kg/$cm^3$]\\
$\mathbf\vec{K} \sim\mathbf\vec{j}$ [kg/$m^2$s] inertial field strength\\ $\mathbf\vec{G}$ [kg/$m^3$]
 gravic field strength\\

Equations (1)- (4) represents a linear partial differential
equation system. The fields are created not by existence but by
changing of mass density and mass flow by space and time.In this
formal sense the gravo-inertial field equations have an analogy in
electromagnetics (Maxwell equations).\\The gravic field 
$\mathbf\vec{G}$
and the gravitational force $\mathbf\vec{F}$ of  a test particle  is
proportional to its mass m. If we imaginate to stop the cosmos
expansion, then the gravic field vanishes. Therefore there is a
gravic field around the particle as result of changed density of
the world by flow and there is a global attraction of all
particles to each other.\\ \begin{equation}
\mathbf\vec{G}(x)=\frac{1}{c^2}\int\limits_0^cd\vec{v}\times\vec{K}\end{equation}\\

 From this force  results the gravitational force (in analogy to Lorentz
force) from the rotational field strength $\mathbf\vec{G}$, which is
proportional to the mass of the particle\\
\begin{equation}
d\mathbf\vec{F}=m\cdot\tau\cdot(d\mathbf\vec{v}\times\mathbf\vec{G})
\end{equation}\\
with\\
$\tau$[$m^4$/skg] as an appropriate constant\\ $\mathbf\vec{F}$
gravitational force\\ $\mathbf\vec{v}$ the velocity of cosmical masses
summed by all masses in all directions\\ $\tau$  a cosmological
time dependent constant.\\

 Because the masses are moved relatively to a particle on x it suffers a force perpendicular to the
velocity $\mathbf\vec{v}$. The system is rotational-symmetric, homogeneous and isotropic.\\
If all masses unmoved, then $\mathbf\vec{G}=0$.\\The
inertial field $\mathbf\vec{K}$ cause a gravitational field $\mathbf\vec{G}$, on
the other hand $\mathbf\vec{G}$ arises $\mathbf\vec{K}$.\\ We can introduce the
common potentials $\varphi$ and $\mathbf\vec{A}$ \\

$\mathbf\vec{K}=-\nabla\varphi-\frac{\mathbf\partial\vec{A}}{\partial t}$\\

 and build the wave equations
\begin{equation}
\nabla^2\varphi-\frac{1}{c^2}\frac{\partial^2 \varphi}{\partial
t^2}=-\frac{\partial\varrho}{\partial t} \end{equation}\\

\begin{equation}
\nabla^2\mathbf\vec{A}-\frac{1}{c^2}\frac{\partial^2\mathbf\vec{A}}{\partial
t^2}=-\frac{\partial\mathbf\vec{j}}{\partial t}
\end{equation}\\

 If the cosmos expands then there is the
energy concentrated in the moved masses and give rise to the term
on the right of equation 1. From the expanding space and time and
the mass flow change by time follows the creation of a field that
contradict the expansion (equation 3). Both fields are connectes
by the equation 3 und 4. The analogy to Maxwell theory is
obviously. In analog way we assume the separation in gravic and
inertial part by four dimensional vector. We obtain a force
action. The Lorentz force correspond  the gravitational force
$\mathbf\vec{F}$ . We recognize the gravitational field as consequequence
of relativiy principle. In this way the gravic field on point x is
a relativistic effect and local transformable. In common sense the
field can only removed by stopping (transforming) of all cosmic
masses in rest reference system. If the cosmos contract then the
gravitation is directed away from the masses. Corresponding
principle of continuity the cosmos is oscillating. The expanding
masses are creating space, time and gravitation. Gravitational and
inertial mass are identically, one is the cause of the other.

\section{Conclusion}
The creation of space and time are connected with fields, only the
masses and the rule of inertia are given. All processes in nature
are determinded by interacting fields or particles, by two energy
forms or in common in twofold way. This state of cosmos is an
instantaneous value which is determined by interaction.
Gravitation is not determined by masses but is in a sense the
result of expanding and creating space and time. In this process
is created inertial field strength $\mathbf\vec{K}$ connected with a
gravitational field strength $\mathbf\vec{G}$.  Because the density change is
cause for creation of gravitational field, the force on particles
is proportional to ist  masses.  The drifting masses produce a
"guiding field" which is with expansion or contraction radial
directed.\\
\begin{figure*}
\caption{Gravo-inertial "coupled" fields}
\medskip\
\centering
\scalebox{0.5}{\includegraphics{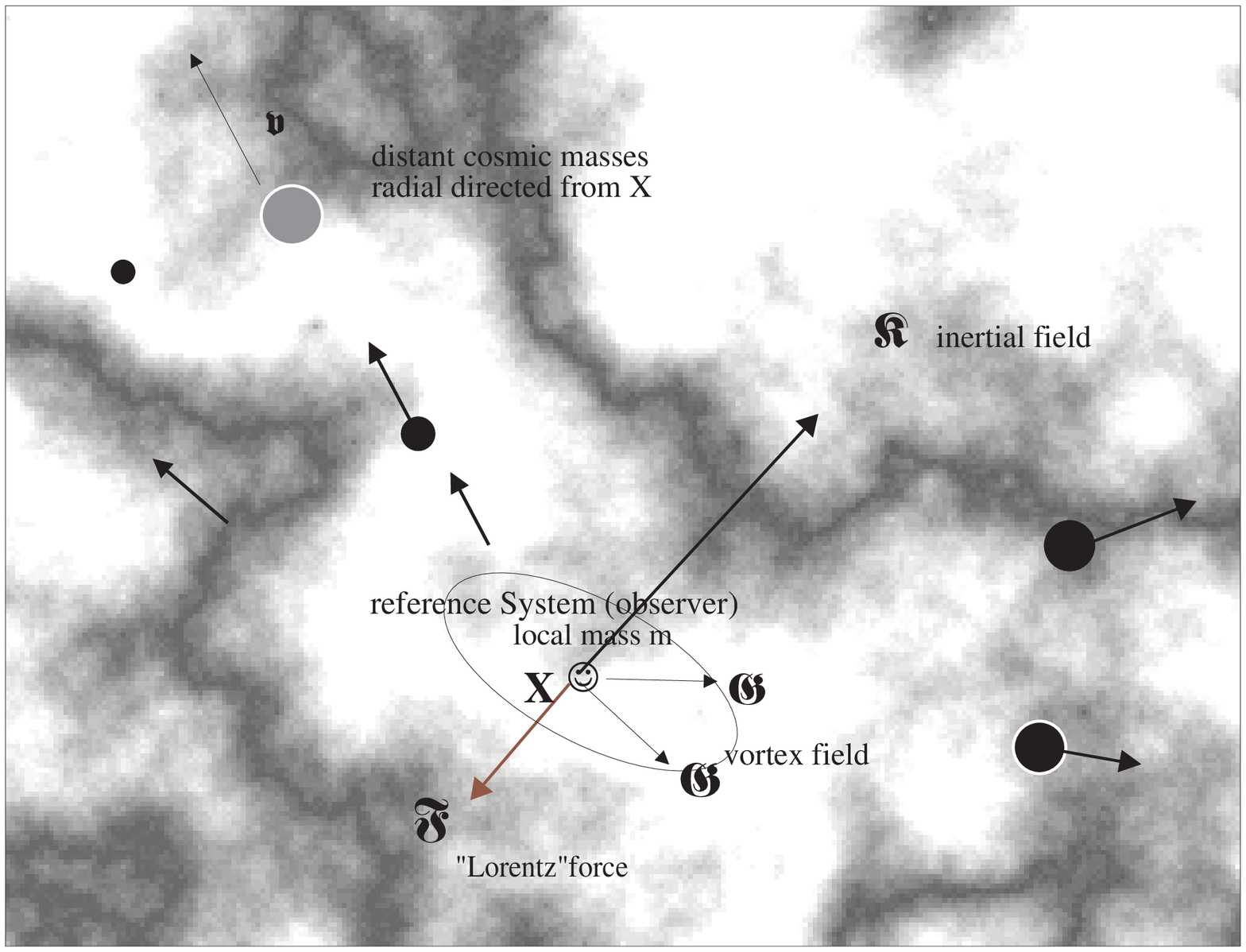}}
\end{figure*}

\end{document}